Elyas Bayati, Raphaël Pestourie, Shane Colburn, Zin Lin, Steven G. Johnson, Arka Majumdar*

# Inverse Designed Extended Depth of Focus Meta-Optics for Broadband Imaging in the Visible

**Abstract:** We report an inverse-designed, high numerical aperture (~0.44), extended depth of focus (EDOF) meta-optic, which exhibit a lens-like point spread function (PSF). The EDOF meta-optic maintains a focusing efficiency comparable to that of a hyperboloid metalens throughout its depth of focus. Exploiting the extended depth of focus and computational post-processing, we demonstrate broadband imaging across the full visible spectrum using a 1 mm, f/1 meta-optic. Unlike other canonical EDOF meta-optics, characterized by phase masks such as a log-asphere or cubic function, our design exhibits a highly invariant PSF across ~290nm optical bandwidth, which leads to significantly improved image quality, as quantified by structural similarity metrics.

## 1 Introduction

Meta-optics are two-dimensional arrays of subwavelength scatterers that are designed to modify different characteristics of light such as its wavefront, polarization, intensity, and spectrum [1-4]. Over the last decade, these devices have enabled ultrathin and flat implementations of various optical elements, including lenses [5-7], vortex beam generators [8-9], holographic plates [10-11], and freeform surfaces [12-14]. Among these elements, metasurface lenses, often termed as metalenses [1, 5-7] have been touted as a promising alternative to conventional, bulky refractive lenses. These metalenses, however, are diffractive and thus exhibit severe chromatic aberrations, limiting their utility and precluding their usage with incoherent, broadband white light. While refractive optical lenses also suffer from material dispersion, the resulting chromatic aberration is significantly lower compared to that of diffractive elements [16], including metalenses, where the severe chromaticity mainly comes from its fixed phase discontinuities at Fresnel zone boundaries [15].

In recent years, several approaches have been proposed to eliminate the chromatic aberrations of metalenses [17-18]. For example, multiplexing different unit cells designed for several discrete wavelengths can enable high-quality polychromatic metalenses [17-18]. In another design, multilayer metasurfaces capable of focusing discrete colors of red, green, and blue have been proposed to alleviate chromatic aberrations [18]. These polychromatic lenses, however, are not suitable for broadband white light imaging. Separately, there have been a number of broadband imaging demonstrations using dispersion engineering [19-23], which employs scatterers that compensate for both the group delay, and group delay dispersion. These broadband metalenses, however, have a constrained design space, limited to small lens aperture and low numerical aperture (NA) due to the limited group delay that is practically achievable with meta-atoms [24-25]. For example, in Ref. 19 [20] the reported achromatic metalenses have a diameter of $200\mu m$ (~$100\mu m$) and a NA of 0.02 (0.106). Moreover, these dispersion-engineered meta-molecules generally entail multiple, coupled scatterers per unit cell, necessitating smaller features relative to standard metalenses, presenting challenges for high-throughput manufacturing [14, 26]. Another approach to mitigate chromatic aberrations is to employ meta-optics in conjunction with computational postprocessing [13, 27-30]. While this method requires additional power and latency due to the computational reconstruction, using appropriately designed meta-optics, broadband images can be captured with large aperture and NA meta-optics, with possible lens diameters in mm-scale to cm-scale. The underlying idea is to extend the depth of focus of the meta-optics, such that all wavelengths reach the sensor in an identical fashion, i.e., the point spread function (PSF) at the sensor plane is wavelength invariant. By characterizing the PSF a priori, we can reconstruct the image from the captured sensor data via deconvolution. This hybrid meta-optical digital approach was first reported using a cubic phase mask (CPM) and Weiner deconvolution [13]. A CPM generates an Airy beam, which propagates through free space without significant distortion. While Airy beams over a broad range of wavelengths behave similarly, the PSF does not resemble that of a lens, which can introduce asymmetric artifacts in images when the signal-to-noise ratio (SNR) is low. Using rotationally symmetric extended depth of focus (EDOF) lenses, the image quality can be improved [31], as recently demonstrated by logarithmic-aspherical [32] and shifted-axicon meta-optics. However, the achievable optical bandwidth is still limited, as the PSF as a function of wavelength exhibits significant variation (as measured by its

**Elyas Bayati and Raphaël Pestourie:** These authors contributed equally to this work.

**\*Corresponding Author: Arka Majumdar,** Department of Electrical and Computer Engineering, University of Washington Seattle, WA-98195, Department of Physics, University of Washington, Seattle, WA 98189, USA, E-mail: arka@uw.edu

**Elyas Bayati and Shane Colburn:** Department of Electrical and Computer Engineering, University of Washington Seattle, WA-98195

**Zin Lin and Steve G, Johnson, Raphaël Pestourie:** Department of Mathematics, MIT, Cambridge, MA 02139, USA,

correlation coefficient [31]). Additionally, many of these EDOF lenses are designed such that segmented regions of the lens focus at different depths, which contributes to haze and a large DC component in the modulation transfer function (MTF).

In this paper, we employ inverse design techniques to create an EDOF meta-optic with broad optical bandwidth while maintaining a lens-like PSF. Here, the design process starts from the desired functionality mathematically described by a figure of merit (FOM), and the scatterers are updated to optimize the FOM. Inverse design techniques have recently generated strong interest in nano-photonics and meta-optics and have been employed to design high efficiency periodic gratings [33, 34], monochromatic lenses [35, 36], PSF-engineered optics [37] and achromatic lenses [38-40] to name a few. Recently, we demonstrated an inverse-designed 1D EDOF cylindrical meta-optic and demonstrated that they outperform existing EDOF metalenses in terms of efficiency while producing a symmetric PSF [41]. In this paper, we report a 2D EDOF meta-optics with four-orders of magnitude larger number of degrees of freedom and demonstrate full-color imaging over a broader optical bandwidth than what is possible using state-of-the-art EDOF metalenses. We show that the optical bandwidth in the inverse-designed meta-optic is 290nm in the visible wavelength range, at least a factor of 2 higher than that of existing EDOF metalenses. Unlike most existing implementations of diffractive or refractive EDOF lenses, the reported EDOF metalens creates a lens-like PSF, without introducing significant blur and maintaining high focusing efficiency.

## 2 Design

We performed a large-scale optimization of meta-optics using the framework reported before [46] and extended to three-dimensional Maxwell's equations [47, 48]. Designing a large-area meta-optic with a brute-force solver is challenging, because meta-optics present two very different length scales: a macro scale (~mm diameter of the whole optic) and a nano scale (the smaller features of the scatterers) which are difficult to resolve concomitantly [49]. To overcome this limit and significantly accelerate the simulation and optimization problems, we use a hybrid multi-scale photonic solver, which relies on a locally periodic approximation [50] where the nanoscale scatterers are approximated by a surrogate model, and the macroscale is solved by a convolution with the analytical free space Green's function [46] to evaluate the field anywhere in the far field.

$$\boldsymbol{E}_{far-field}(\boldsymbol{x}) = -\int_{metasurface} \boldsymbol{G}(\boldsymbol{x}, \boldsymbol{x}') \cdot \boldsymbol{E}_{local}(\boldsymbol{x}'|ps) d\boldsymbol{x}',$$

where **G** are the appropriate Green's functions and $\boldsymbol{E}_{local}$ is defined piece-wise constant by evaluating the transmission through the unit cell for each of the parameters $ps$ of the metasurface [48]. In the present paper, we fit a surrogate model that quickly evaluates the complex transmission of a unit cell as a function of its parameter change using Chebyshev interpolation [51]. We considered two parameterizations of the unit-cell geometry with a period of 400 nm: for the 100-μm-focal-length meta-optic, a cylindrical pillar of silicon nitride (SiN) on top of silica with a diameter varying from 100 nm to 300 nm (considering a minimum fabricable feature of 100 nm); for the 1-mm-focal-length meta-optic, a square nanofin of SiN on top of silica with each side of the nanofin varying from 100 to 300 nm. When the number of parameters for the scatterer is greater than five, Chebyshev interpolation suffers the "curse of dimensionality", and a neural network surrogate becomes preferable [52]. The gradient of the field computed using this approximate solver is obtained using an adjoint method [48]

$$\nabla_{ps}\boldsymbol{E}_{far-field}(\boldsymbol{x}) = -\int_{metasurface} \boldsymbol{G}(\boldsymbol{x}, \boldsymbol{x}') \cdot \nabla_{ps}\boldsymbol{E}_{local}(\boldsymbol{x}'|ps) d\boldsymbol{x}',$$

where $\nabla_{ps}$ denotes the gradient with respect to the metasurface parameters. The figure of merit to be maximized is the minimum intensity along the focal axis [41] and the depth of field was designed for an incident plane wave at 633 nm

$$maximize \left(\min_{p \in P} I(p)\right),$$

where $P$ is a set of target points along the focal axis, and $I(p) = \bar{E}_{far-field}(p) E_{far-field}(p)$ is the intensity computed at 633 nm. The ¯ notation denotes the complex conjugate. We used the equivalent epigraph form to make the optimization formulation differentiable [47]. We found that a spacing of 1.2 μm between two evaluations of intensity along the optical axis is optimal to result in a flat intensity profile over the desired depth of field. Consequently, in contrast to the depth of focus of a traditional lens, which depends on the ratio between focal length of the device and its diameter, the number of intensity evaluations for the EDOF minimax problem is proportional to the diameter of the device. For the 100-μm-focal-length meta-optic, we maximized the minimum intensity over 65 evaluations between a depth of 75 μm and 150 μm. For the 1-mm-focal-length meta-optic, we maximized the minimum intensity over 422 evaluations between a depth of 0.75 mm and 1.25 mm. Note that the latter optimization has 6.5 times more points in the set $P$ than the former, therefore resulting in a poorer optimum. To ensure polarization insensitivity, the designed meta-optic has C4v symmetry. The figure of merit was optimized using the "CCSA-MMA" algorithm [59], which is a gradient-based optimization algorithm that can handle non-linear constraints and ensures to converge to a local optimum. We also designed two traditional metalenses with the same focal lengths of the inverse-designed ones using traditional forward design method [12]. This forward design technique is based on rigorous coupled-wave analysis (RCWA) [42] and finite-difference time-domain (FDTD) simulation. In the forward design method, we arrange the scatterers on a subwavelength lattice, and spatially vary their dimensions based on the spatial phase profile of a hyperboloid metalens [7]. We note that the reported meta-optics is one of the largest inverse-designed nanophotonic structure reported in the literature.

To validate our design, we fabricated both the inverse-designed EDOF meta-optics and forward-designed traditional hyperboloid metalens. The diameter for these metalenses is 1mm and 100 μm. The EDOF and traditional metalenses are fabricated in silicon nitride on the same sample. We first deposited 600-nm-thick layer of silicon nitride using Plasma-enhanced chemical vapor deposition (PECVD) on a 500- μm -thick fused-silica. The sample was spun coated with a layer of 200-nm-thick positive electron beam resist (ZEP-520A), followed by an additional layer of conductive material (Au/Pd) to avoid charging effects during electron beam lithography. After that, the pattern is written using electron-beam lithography (JEOL JBX6300FS at 100 kV), and the exposed sample was developed in amyl acetate. Next, around 50-nm-thick aluminum was evaporated directly onto the developed sample. After performing lift-off, the silicon nitride layer is etched through its thickness using an inductively coupled plasma etcher with a mixture of CHF3 and O2 gases. In the end, the remaining aluminum was removed in AD-10 photoresist developer.

Figs. 1A-B show the scanning electron micrographs (SEMs) of the fabricated inverse-designed EDOF meta-optics and a zoom-in SEM showing the silicon nitride nano pillars forming the circularly symmetric EDOF meta-optics. A custom microscope setup (see Section 5.1 and Figure 5) under illumination by a 625 nm light-emitting diode (part number Thorlabs-M625F2) is used to measure the depth of focus of our fabricated meta-optics. Figs. 1C-F show the experimentally measured intensity profiles along the optical axis for the two traditional metalenses and two inverse-designed EDOF meta-optics. The intensity profiles along the optical axis are captured using a camera and translating the microscope along the optical axis using an automated translation stage with an axial resolution of 1 μm (Figs. 1C, E) and 2 μm (Figs. 1D, F). We clearly observe an elongation of the focal spot along the optical axis. This behavior can be clearly explained by showing the PSF at two different depths (1 mm and 1.09 mm) for our 1 mm aperture lenses for both EDOF and traditional metalens as shown in Figs. 1G-J. We identified the depth of focus as the range along the optical axis, where the beam profile remains Gaussian. We measured the depth of focus to be ∼83 μm (∼0.4 mm) for our 100 μm (1 mm) EDOF meta-optics. The depth of focus is only 10 μm (0.1mm) for the 100 μm (1mm) aperture traditional metalens. We also measured the focusing efficiency of the 1 mm aperture lenses along the optical axes (see Section 5.2 and Figure 6). We did not observe significant degradation of the efficiency in the designed EDOF optics (efficiency of 47%) compared to the traditional metalens (efficiency of 51%). We note that, the center focal length of the EDOF lens is slightly different from the designed ones, which are same as the traditional metalens. We attribute this to the fabrication imperfections, which may affect the performance of the EDOF metalens more than traditional metalens. However, due to the EDOF nature, the designed focal length is within the depth of focus.

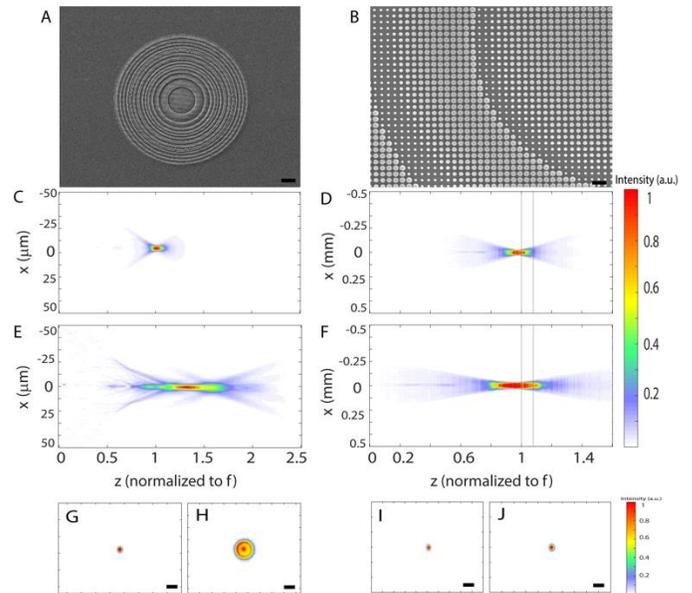

Figure 1. Scanning electron micrograph (SEM) of (A) inverse-designed EDOF meta-optics, (B) Zoom-in SEM on inverse-designed EDOF meta-optics showing silicon nitride scatterers forming the EDOF metalens; The scale bars correspond to 10 μm (A) and 1 μm (B), (C-F) Experimentally measured field profile along optical axis for traditional (C, D) and inverse-designed EDOF metalenses (E, F) with focal lengths of (C, E) $f = 100\ \mu m$, (D, F) $f = 1\ mm$, (G-J) The PSFs of 1mm aperture traditional metalens (G, H) and EDOF metalens (I, J) at two different depths correspond to (D, F). Scale bars, 50 μm.

Fig. 2A. shows the cross-sections of the focal plane of the 1 mm and 100 μm aperture inverse designed EDOF and 1 mm traditional metalens to compare the PSFs. We also fit the intensities near the focal plane using a Gaussian function to estimate the full width at half maximum (FWHM) of the beam profiles. The minimum FWHM for the fabricated EDOF meta-optics and traditional metalens with 1 mm (100 μm) aperture focal length is 24.9 μm (3.35 μm) and 20.6 μm (3μm), respectively as shown in Fig. 2A. The diffraction-limited spot of our 1mm aperture metalens with the same geometric parameters is 1.46 μm. We note that, in this paper, the diffraction limited FWHM for a lens is calculated by fitting a Gaussian function to the Airy disk with the same geometric parameters, as shown in Ref [7]. The difference between the diffraction limit and the experimentally measured FWHM comes from the geometric aberration. This aberration is expected to scale linearly with the aperture [53]. Hence, the FWHM for 1mm lens is almost ∼10 times larger than that of the 100μm lens, even though both aperture lenses have the same NA. We note that, these aberrations could be further minimized by redefining the FOM in the inverse design. Nevertheless, the PSF for both EDOF and traditional lenses are quite similar, unlike traditional EDOF lenses [31]. The PSF of our EDOF metalens, however, remains identical over an extended distance along the optical axis, making the PSF insensitive to focal shift, e.g., due to the wavelength change. This property enables the EDOF meta-optics to produce similar PSFs for all wavelengths at one specific plane. On the other hand, the PSF of a standard metalens varies significantly with wavelength.

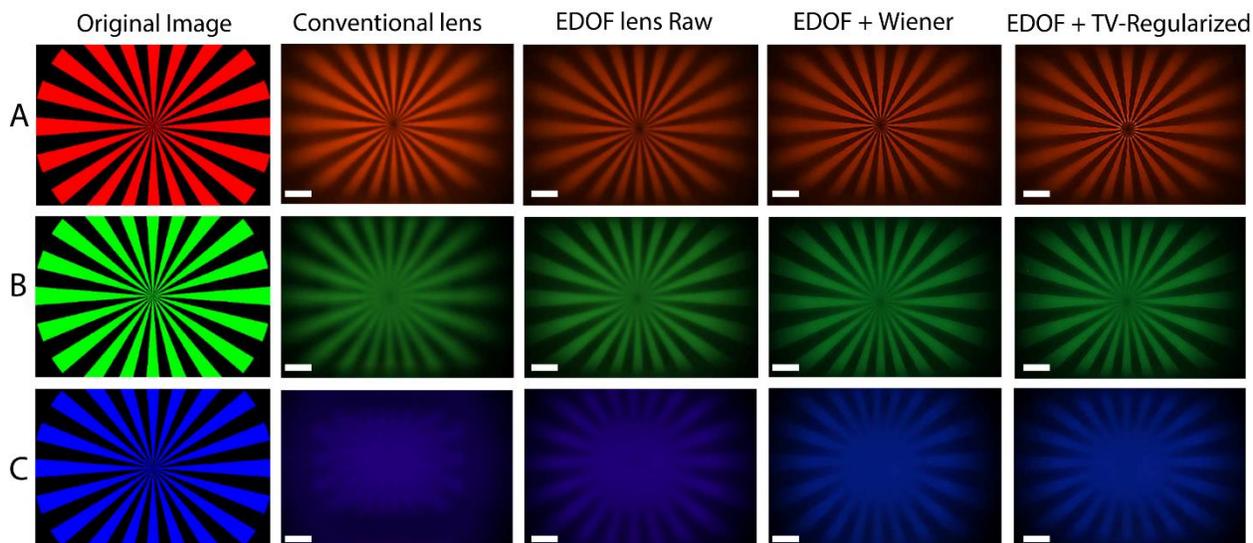

Figure 3. **Imaging at discrete wavelengths**. The appropriately cropped original object patterns used for imaging are shown in left side of the figure. Images were captured with the singlet traditional metalens, the EDOF lens without, and with deconvolution (EDOF + Wiener), and with the total variation (TV) regularization. Images were captured using narrowband (bandwidth of 20nm) LEDs with center wavelengths at (A) 625 nm, (B) 530 nm and (C) 455 nm wavelength. The scale bars correspond to 100 μm.

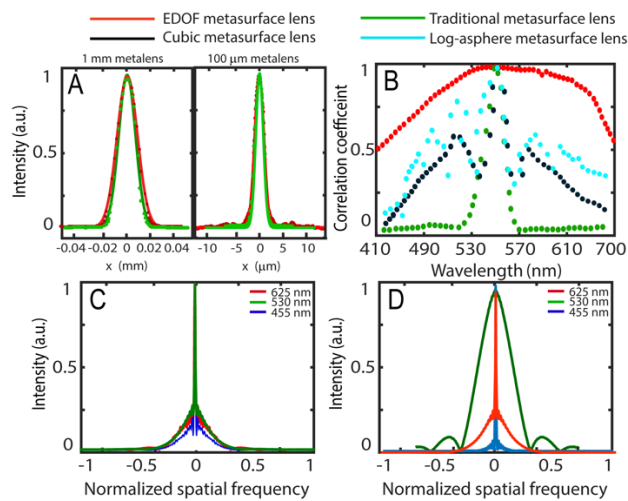

Figure 2. **Characterization of the metasurfaces:** (A) PSF of the traditional and EDOF metalenses for 1mm and 100 μm focal length. (B) Correlation coefficient plots for four different metalenses as a function of wavelength. (C-D) The corresponding x–y plane cross sections of the experimental MTFs for our EDOF metalens (C) and traditional metalens (D) are displayed with red lines from its PSF measured under red light, green lines under green light, and blue lines under blue light. The MTF plots have a spatial frequency normalized to 560 cycles/mm.

To quantitatively understand this behavior, the MTF is calculated for our EDOF and a traditional metalens as shown in Figs. 2C, D. The MTF of a traditional lens measured at the focal plane for 530 nm preserves more spatial frequency information for green light (530 nm) compared to red (625 nm) and blue (455 nm). On the other hand, the MTF for our EDOF meta-optics remains same for all three colors. In fact, not just these three colors, the MTF of the EDOF meta-optics remains same over the whole visible range. Additionally, the EDOF meta-optics also exhibits a higher cutoff frequency than a traditional metalens

Having a highly wavelength-invariant MTF across a large optical bandwidth is necessary for applying computationally efficient deconvolution techniques to recover high-quality, full-color images. One metric for quantifying this invariant behavior is to calculate the correlation between PSFs. The correlation can be calculated as the inner product between a reference PSF (here we used the PSF at 530 nm) and the PSFs at all other wavelengths in the range of interest. To compare the performance of our inverse designed EDOF metalens with other existing EDOF metalenses, we also calculate the correlation coefficients of log-asphere and cubic metalenses [31] (Fig. 2B). To make a fair comparison, all these meta-optics are designed at 530 nm wavelength. Fig. 2B clearly shows that the correlation coefficient of our inverse designed EDOF is higher and far more uniform compared to other metalenses over a broad wavelength range. We defined the optical bandwidth as the range of optical wavelengths where this correlation coefficient is greater than 0.5 [43]. We estimated the optical bandwidths relative to a central wavelength (530 nm) to be 290 nm (inverse-designed EDOF), 130 nm (log-asphere metalens), 95 nm (cubic metalens), and 15 nm for a standard metalens. We also calculated the fundamental limit on the achievable optical bandwidth based on the given thickness and numerical aperture of our meta-optics [54]. This fundamental limit is only ~20 nm for a hyperboloid metalens, consistent with our experimental results and significantly lower than that achieved with our EDOF designs.

## 3 Imaging Results

A larger aperture lens collects more photons and thus can capture images with higher SNR. Consequently, we focus on 1 mm EDOF meta-optics for imaging. The captured image in our system can be written in the form of $f = Kx + n$ [44], where $K$ is the system kernel or PSF, $x$ is the desired image, and $n$ is the noise that corrupts the captured image $f$. While there are several methods to estimate the image $x$, we chose to use Wiener and total variation (TV)-regularized deconvolution due to their effectiveness and generalizability to different scenes relative to deep learning-based deconvolution methods.

Our system kernel is sampled by measuring PSFs for our EDOF metalens at three wavelengths using separate LED sources centered at 625 nm, 530 nm and 455 nm (Thorlabs M625F2, M530F2, M455F1). To demonstrate the imaging capability of our EDOF system, we first considered narrowband LED imaging (bandwidth ~20nm) for three different wavelengths. Fig. 3 shows our EDOF imaging performance compared to a traditional metalens with and without deconvolution. Although our original image is focused for red light as shown in Fig. 3A, the images captured under green and blue light (Fig. 3B, C) are severely distorted due to strongly defocused PSFs in these wavelengths. On the other hand, the images which are captured by inverse-designed EDOF metalens under green and blue illumination (even before deconvolution) are less blurry compared to those of a standard metalens, thanks to their symmetric and wavelength-invariant PSFs. After deconvolution, the captured images appear even sharper and in focus for all wavelengths. We noticed that our image under blue light illumination is not as focused as red and green. Although our figure of merit includes the blue wavelength in the inverse design method, we attribute the lower image quality for blue to its PSF correlation coefficient being lower compared to other part of the visible regime. To quantitatively estimate the imaging quality, we calculated the structural similarity index measure (SSIM) [56] for the captured images in Fig. 3. The calculated SSIMs between the original image and the output images from EDOF (traditional) metalenses are 0.637 (0.492) for the red channel, 0.503 (0.239) for the green channel, and 0.363 (0.149) for the blue channel. Thus, for all the wavelengths, our EDOF metalens provides a higher SSIM than that of the conventional metalens. In addition to Wiener deconvolution algorithm, we used TV-regularization [45] for image denoising and deconvolution. TV regularization is based on the statistical fact that natural images are locally smooth, and the pixel intensity gradually varies in most regions.

Finally, our EDOF system was tested with a full-color OLED display (SmallHD 5.5 in). The OLED monitor was placed ~13 cm away from our EDOF metalens, which displays some of our ground-truth images as shown in the Fig. 4. The images were chosen in a way to have all the combination of visible colors. The chromatic aberration is clearly observed on captured images with a conventional metalens. However, the raw images which are captured by the EDOF metalens show less chromatic aberration, even though they are still blurred before deconvolution. We then computationally processed the captured images using Wiener deconvolution and TV regularization algorithm. The post processed images show in focus, smooth and less aberrated images. We note that such deconvolution cannot mitigate the chromatic aberration in the conventional metalens as the MTFs outside green wavelength is too distorted. We also noticed that, for the case of colorful floating balloons (Fig. 4D), the chromatic blur for blue balloons is still there, due to the low PSF correlation coefficient in this wavelength as explained earlier. Nevertheless, the achromatic performance of the inverse-designed meta-lens is significantly better than that of the metalens. We note that due to the narrow MTF of the metalens over a broad wavelength range, any deconvolution operation on the post-capture images significantly amplifies the noise, making the reconstructed images significantly worse than the captured ones. Hence, we only report the captured data for a metalens, and no reconstructed images are shown here.

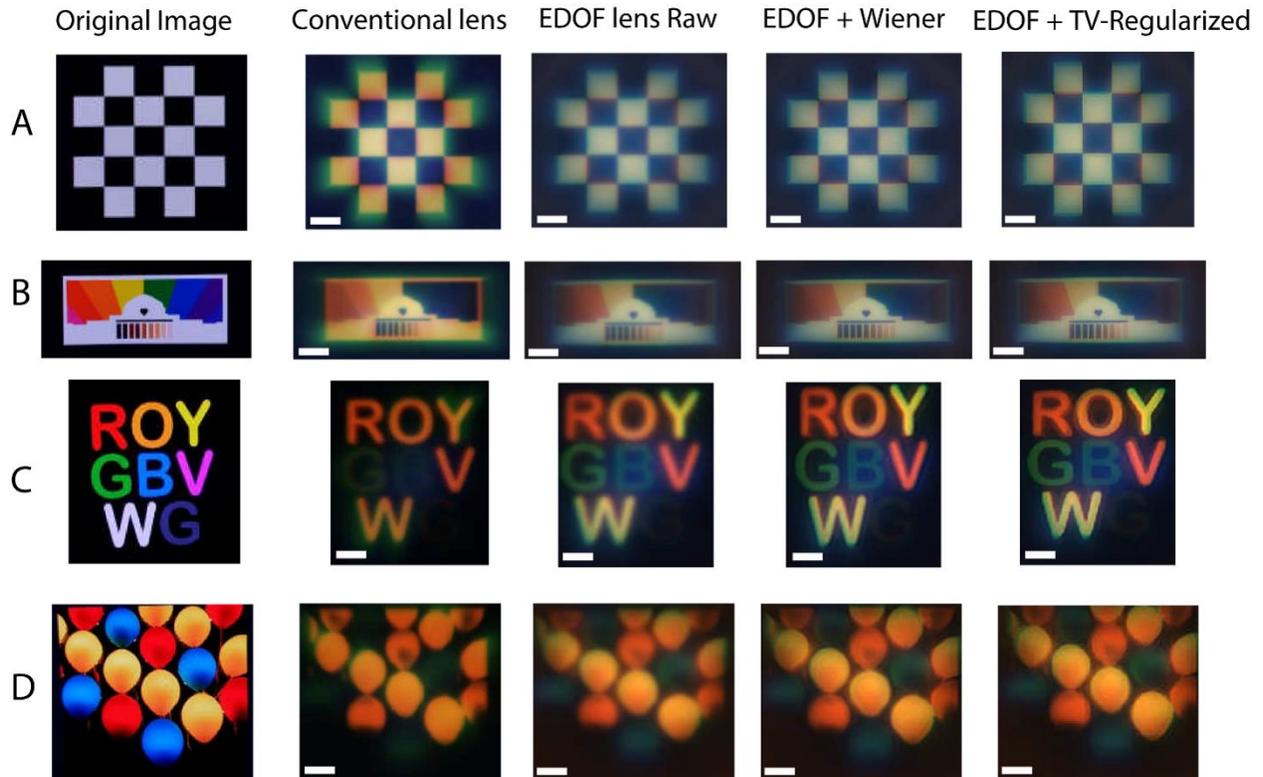

Figure 4. **White light imaging**. Images were taken from an OLED display of (A) white-black cross pattern, (B) colored rainbow MIT pattern, (C) colored letters in ROYGBVWG and (D) colorful floating balloons. The appropriately cropped original object patterns used for imaging are shown in the left column. The scale bars correspond to 100 μm.

We also characterized the imaging quality of our inverse designed EDOF meta-optics compared to other existing EDOF metalenses such as cubic or log-aspheres, by calculating the SSIM for "ROYGBVWG" image (see Section 5.3, Figure 7 and Table 1). The calculated SSIMs between the original image and the output images from inverse designed EDOF metalenses are higher than the SSIM values from other EDOF lenses (see, Table 1). Thus, our inverse designed EDOF metalens indeed provide a higher SSIM score and better imaging quality than other existing EDOF lenses. This already shows the benefit of inverse design, which can be further improved by considering angle-dependence of the scatterers and other aberrations in the FOM.

## 4 Discussion and Conclusion

Using an inverse design method, for the first time, we created a 2D EDOF metalens that has a symmetric PSF, large spectral bandwidth, and that also maintains focusing efficiency comparable to that of a traditional metalens. The proposed inverse designed EDOF meta-optics in conjunction with computational imaging is used to mitigate the chromatic aberrations of traditional metalenses. Unlike other existing EDOF metalenses (log-asphere lenses or other variants), our inverse-designed EDOF metalens represents a highly invariant PSF across a large spectral bandwidth to improve the full-color image quality. We have quantified this via calculating and comparing the PSF correlation coefficient with other EDOF metalenses over the visible spectral range. We emphasize that this is the largest aperture (1 mm) meta-optics ever reported for achromatic imaging, and as such there is no fundamental limitation on increasing aperture, like dispersion engineered meta-optics do.

While we extended the spectral bandwidth and demonstrated achromatic imaging even without post processing algorithms, our images still need to be computationally processed to give us in-focus and better-quality images. This postprocessing adds latency and power to the imaging system. Additionally, we have primarily focused on chromatic aberration corrections, and Seidel aberrations are largely ignored. Using a more involved figure of merit and including the angular dependence, a longer depth of focus and larger field of view metalens can be designed, which can potentially preclude the use of computational postprocessing. In this work, we also emphasize the utility of rotationally symmetric EDOF metalenses (for simplified packaging and mitigating deconvolution artifacts), as well as larger apertures (1 mm) compared to our previous works. A larger aperture will enable more light collection, higher signal-to-noise ratio and faster shutter speed, which are crucial for practical applications. We envision that the image quality can be further improved by co-designing the meta-optics and computational back end [57, 58], as has already been reported to provide very high-quality imaging [57]. Finally, the EDOF metalens has lower volume under MTF curve compared to a metalens at the design wavelength (here green). However, by sacrificing this MTF in one wavelength, we get back uniformity of the MTF over a broader wavelength range. This suggests a fundamental trade-off between the optical bandwidth and achievable MTFs, which to the best of our knowledge is not studied before and will constitute a future theoretical study.

# 5 Methods

## 5.1 Confocal microscopy setup

A confocal microscopy setup under illumination by a 625 nm light-emitting diode (part number Thorlabs-M625F2) is used to measure extension of depth of focus for our fabricated metalenses. Fig. 5. shows the confocal microscopy setup that we have used to experimentally measured field profiles along the optical axis for the two traditional metalenses and two inverse-designed EDOF metalenses.

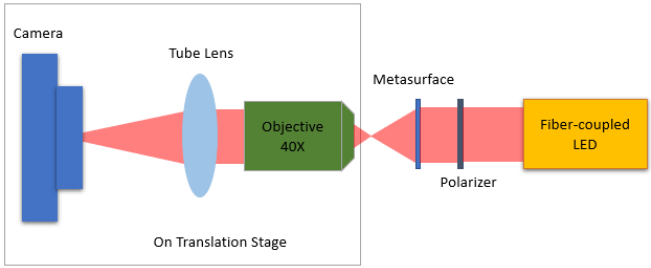

Figure 5. Confocal microscopy setup used to measure the metalenses.

## 5.2 Experimental performance of the metalens

We define the focusing efficiency as the power within a circle with a radius of three times the minimum full-width half maxima (FWHM) at the focal plane to the total power incident upon the metalens [7]. We plot the focusing efficiency of the metalenses along the optical axis (Fig. 6). We expect the focusing efficiency to remain the same along the depth of the focus (for our EDOF metalenses), and then drop off as we longitudinally move away from the depth of focus. Clearly, for the EDOF metalens, the efficiency remains high over a longer depth as expected. We did not observe significant degradation of the efficiency in the designed EDOF lenses compared to that of an ordinary lens.

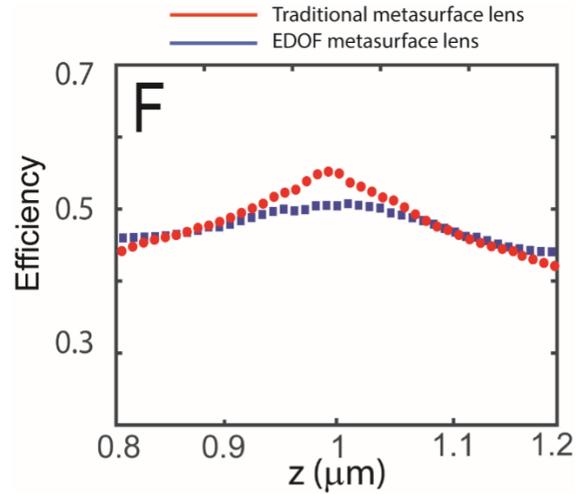

Figure 6. Experimentally Measured focusing efficiency of 1mm EDOF and traditional metalenses as a function of distance along the optical axis.

## 5.3 Comparison with other potential EDOF lenses in terms of SSIM and imaging quality

In order to see how imaging results of our inverse-designed EDOF lens outperform the standard and other potential EDOF lenses, we also characterized the imaging quality of our inverse designed EDOF metalens compared to other existing EDOF metalenses such as cubic functions or log-aspheres, by calculating the SSIM for "ROYGBVWG" image. The SSIM calculations are sensitive to the subject's translation, scaling, and rotation, which are difficult to eliminate in an experimental setup. Hence, we utilize the simulated PSFs and phase functions (cubic functions and log-aspheres) with the same geometric parameters to perform imaging and post processing for our "ROYGBVWG" image as it is shown in Fig. 7. The calculated SSIMs between the original image and the output images from inverse designed EDOF, conventional, cubic and log-asphere metalenses are shown in Table. 1. for the red, green, and blue channel. Clearly, our inverse designed EDOF metalens provides a higher SSIM score and better imaging quality than other existing EDOF lenses.

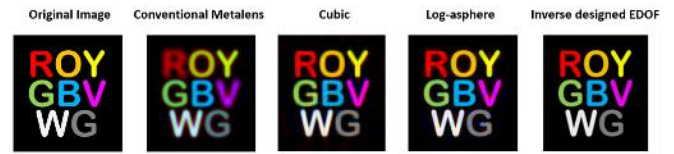

Figure 7. Simulated imaging performance after deconvolution. Deconvolved images captured by the EDOF imaging system, using the simulated images and PSFs.

**Table 1. SSIM on Simulation-Restored Images for standard and different EDOF metalenses**

| Color channel | Inverse designed EDOF | Conventional Metalens | Cubic Metalens | Log-asphere Metalens |
|---|---|---|---|---|
| Red | 0.639 | 0.292 | 0.3062 | 0.4537 |
| Green | 0.7903 | 0.4691 | 0.6543 | 0.7170 |
| Blue | 0.4863 | 0.3449 | 0.4776 | 0.5718 |

**Funding.** The research is supported by NSF-1825308 and DARPA (Contract no. 140D0420C0060). A.M. is also supported by a Washington Research Foundation distinguished investigator award. R. P. is supported in part by IBM Research, the MIT-IBM Watson AI Laboratory, and the U.S. Army Research Office through the Institute for Soldier Nanotechnologies (under award W911NF13-D-0001). Part of this work was conducted at the Washington Nanofabrication Facility/ Molecular Analysis Facility, a National Nanotechnology Coordinated Infrastructure (NNCI) site at the University of Washington with partial support from the National Science Foundation via awards NNCI-2025489 and NNCI-1542101.

**Disclosures.** The authors declare that they have no conflict of interest.